\documentclass[reprint,raggedbottom,aps,physrev,amsfonts,amsmath,amssymb,floatfix,superscriptaddress]{revtex4-2}

\usepackage{graphicx,hyperref, lineno} 
\usepackage[all]{hypcap} 
\hypersetup{colorlinks=true,linkcolor=blue,citecolor=blue,urlcolor=blue}
\usepackage{siunitx}

\usepackage[caption=false]{subfig}
\begin{document}

\title{Atomic-Level Features for Kinetic Monte Carlo Models of Complex Chemistry from Molecular Dynamics Simulations}
\author{Vincent Dufour-D\'{e}cieux}
\email{vdufourd@stanford.edu}
\affiliation{Department of Materials Science and Engineering, Stanford University, Stanford, CA 94305, USA}
\author{Rodrigo Freitas}
\affiliation{Department of Materials Science and Engineering, Stanford University, Stanford, CA 94305, USA}
\affiliation{Department of Materials Science and Engineering, Massachusetts Institute of Technology, Cambridge, MA 02139, USA}
\author{Evan J. Reed}
\email{evanreed@stanford.edu}
\affiliation{Department of Materials Science and Engineering, Stanford University, Stanford, CA 94305, USA}

\date{\today}

\begin{abstract}
\begin{center}
    \textbf{Abstract}
\end{center}
 The high computational cost of evaluating atomic interactions recently motivated the development of computationally inexpensive kinetic models, which can be parametrized from MD simulations of complex chemistry of thousands of species or other processes and accelerate the prediction of the chemical evolution by up to four order of magnitude. Such models go beyond the commonly employed potential energy surface fitting methods in that they are aimed purely at describing kinetic effects.  So far, such kinetic models utilize molecular descriptions of reactions and have been constrained to only reproduce molecules previously observed in MD simulations.  Therefore, these descriptions fail to predict the reactivity of unobserved molecules, for example in the case of large molecules or solids. Here we propose a new approach for the extraction of reaction mechanisms and reaction rates from MD simulations, namely the use of atomic-level features. Using the complex chemical network of hydrocarbon pyrolysis as example, it is demonstrated that kinetic models built using atomic features are able to explore chemical reaction pathways never observed in the MD simulations used to parametrize them, a critical feature to describe rare events. Atomic-level features are shown to construct reaction mechanisms and estimate reaction rates of unknown molecular species from elementary atomic events. Through comparisons of the model ability to extrapolate to longer simulation timescales and different chemical compositions than the ones used for parameterization, it is demonstrated that kinetic models employing atomic features retain the same level of accuracy and transferability as the use of features based on molecular species, while being more compact and parametrized with less data.  We also find that atomic features can better describe the formation of large molecules enabling the simultaneous description of small molecules and condensed phases.
\end{abstract}

\maketitle

\section{Introduction}
Understanding mechanisms of chemical reaction kinetics is a challenging task. Reaction networks can be enormous and complex, even for simple chemical systems containing no more than two or three species. Yet, much effort has been put in advancing our knowledge of chemical kinetics due to its important role in several fields of science, such as fuel combustion \cite{harper2011combustionmechanism, westbrook2009combustionmechanism}, astrophysics \cite{ross1981diammonduranus, kraus2017formation}, polymer science \cite{codari2012polymer}, and organic chemistry \cite{pande2014abinitionanoreactor, meisner2019originoflife}. Atomistic simulations have long been a valuable tool for reaction-mechanism discovery as they complement many of the experimental efforts \cite{harper2011combustionmechanism, westbrook2009combustionmechanism} by providing unrestricted access to the contributions of each atom along specific reaction pathways. This level of detail is difficult to be achieved through experimental methods alone.

\begin{figure*}[htb]
  \centering
  \includegraphics[width=0.9\textwidth]{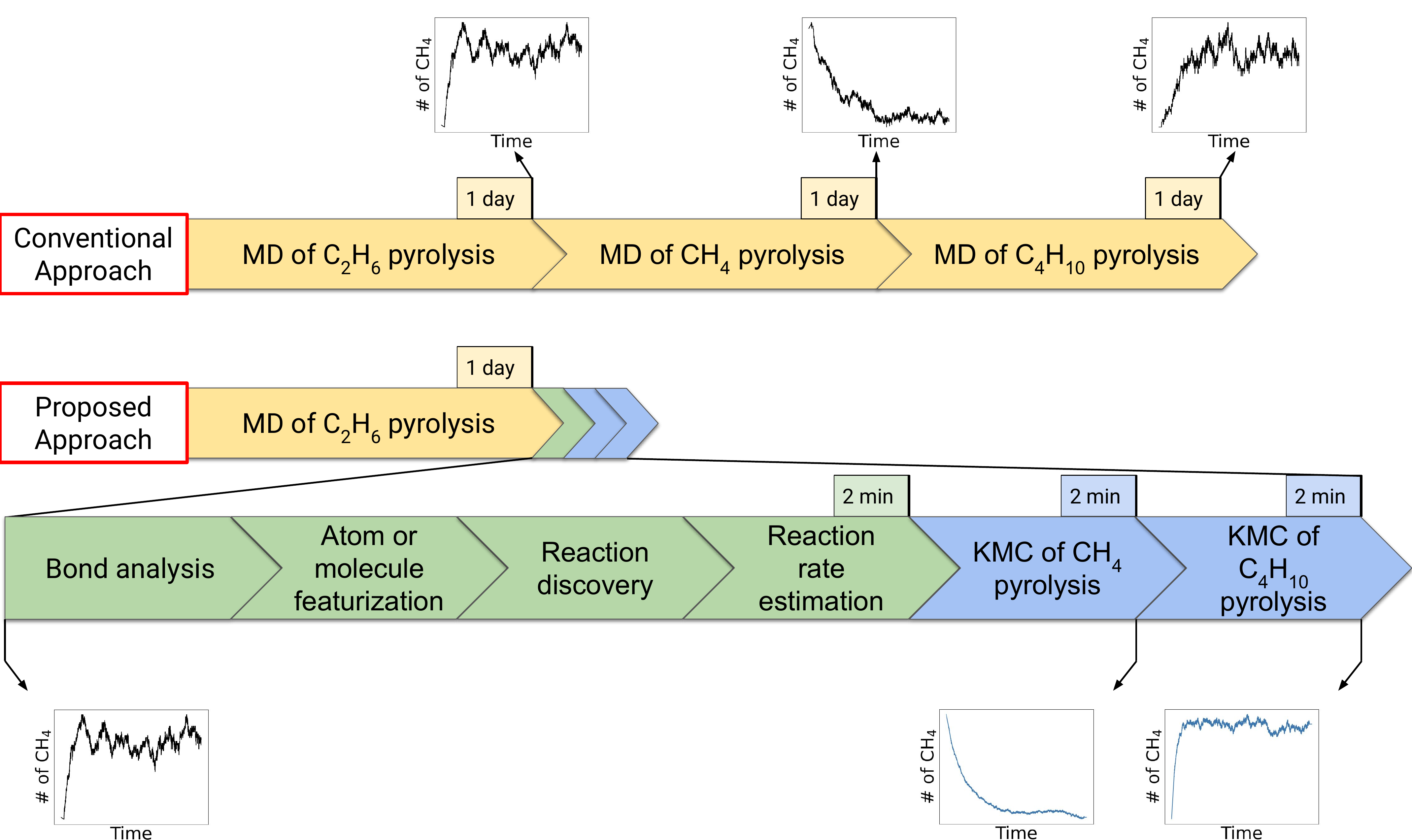}
  \caption{Illustration of two different approaches to obtain the chemical reaction kinetics of three different systems with starting compositions consisting of only one type of molecule: CH$_4$, C$_2$H$_6$, or C$_4$H$_{10}$. In the ``Conventional Approach'' one MD simulation is performed for each of the three compositions. The ``Conventional Approach'' is time-consuming but accurate. In the ``Proposed Approach'', first developed by \citet{EnzeArticle}, a single MD simulation is performed for one of the three compositions. From the data of this MD simulation the observed reaction mechanisms and reaction rates are extracted. This information is in turn employed to obtain the chemical reaction kinetics of the two remaining systems through the use of a computationally inexpensive kinetic model (namely Kinetic Monte Carlo (KMC) simulations). The ``Proposed Approach'' is faster than the ``Conventional Approach'' and can be made just as accurate by the judicious choice of the kinetic model. The simulation times in the illustration are representative of the simulations performed in this work. They are presented only to provide a sense of the computational speedup provided by the ``Proposed Approach''.}
  \label{fig:simulation_approach}
\end{figure*}

One of the main limitations of atomistic simulations is their time scale. Such simulations are often restricted to events that occur in the range of femtoseconds to hundreds of nanoseconds due to the intrinsic need to resolve processes step by step. Several fruitful approaches have been developed with the goal of overcoming these limitations, including parallel replica \cite{joshi2013parallelreplica, voter1998parallelreplica}, bias potential \cite{voter1997biaspotential, hamelberg2004biaspotential}, enhanced sampling \cite{pan2016enhancedsampling}, GPU computing \cite{liu2008GPUMD}, and transition-path sampling \cite{pande2014abinitionanoreactor}. From these advances it became clear that no single method would solve all the timescale limitations of atomistic simulations. Instead, each approach has a niche of applications for which it is best suited for. Futhermore, ease of use by non-experts in commonly used codes is a generally desirable feature of an algorithm that can significantly enhance adoption and use by a much broader spectrum of researchers, e.g. many density functional theory codes. Here, we focus on such an approach where atomic simulation methods are used to automatically parametrize kinetic models of chemical reactivity that naturally give access to extended time scales (Fig.~\ref{fig:simulation_approach}). This algorithm takes as input MD simulations of the type that are already routinely performed by many researchers, and squeezes additional capabilities out of those simulations at negligible cost compared with performing the MD simulations.  Our specific application in this work is in capturing and understanding the chemical reactivity of hydrocarbons.

Atomic-level simulations have contributed much to the understanding of the mechanisms of pyrolysis and combustion of hydrocarbons \cite{cheng2012reaxffmechanismhydrocarbon, dontgen2015automated, he2014reaxffmechanismhydrocarbon, yan2013reaxffmechanismhydrocarbon, lummen2010reaxffcombustion, liu2011reaxffhydrocarbonmechanism, bai2012reaxffhydrocarbonmechanism, liu2014reaxffhydrocarboncombustion, cheng2014reaxffhydrogencombustion, zeng2020reacnetgenerator, liang2020skeletal}. Reaction mechanisms and reaction rates can be extracted \cite{cheng2012reaxffmechanismhydrocarbon, dontgen2015automated, he2014reaxffmechanismhydrocarbon}
from atomistic simulations such as MD simulations and compared directly to experimental data. Alternatively, this information can also be employed to parametrize a kinetic Monte Carlo (KMC) \cite{dontgen2015automated, QianArticle, QianArticle2016, QianArticle2019, QianArticle2020, YanzeArticle} model that can be used to reproduce chemical kinetics for longer time scales at a much reduced computational cost, as illustrated in Fig.~\ref{fig:simulation_approach}.

These kinetic model extractions characterized each reaction by the molecules involved. This approach showed good accuracy in reproducing initial MD simulations evolution \cite{QianArticle, EnzeArticle}, but it has several disadvantages. One of the main disadvantages of employing mechanisms and rates of reactions obtained in MD simulations described in terms of molecules (e.g. A+B$\rightarrow$AB where A, B and AB are molecular species) to parametrize KMC models is that the resulting KMC simulations are bound to only create molecules that have been previously observed in the atomistic simulation. If there exists a molecule that takes longer to be created than the accessible time scale of the MD simulation, then the KMC simulation will not be able to create that molecule either, despite being able to simulate chemical kinetics for longer time scales than MD simulations. Consider for example the process of creation and growth of soot particles. Such particles are the result of incomplete combustion of hydrocarbons and are composed of long carbon chains. Small carbon chains grow in size by aggregation of other molecules. Any MD simulation can only study the growth of carbon chains up to a certain length due to the time scale limitation. Because larger carbon molecules are not observed in the MD simulation, a KMC model would not be able to predict the growth of carbon molecules beyond that specific size observed in the MD simulation.

\begin{figure*}[htb]
  \centering
  \includegraphics[width=0.9\textwidth]{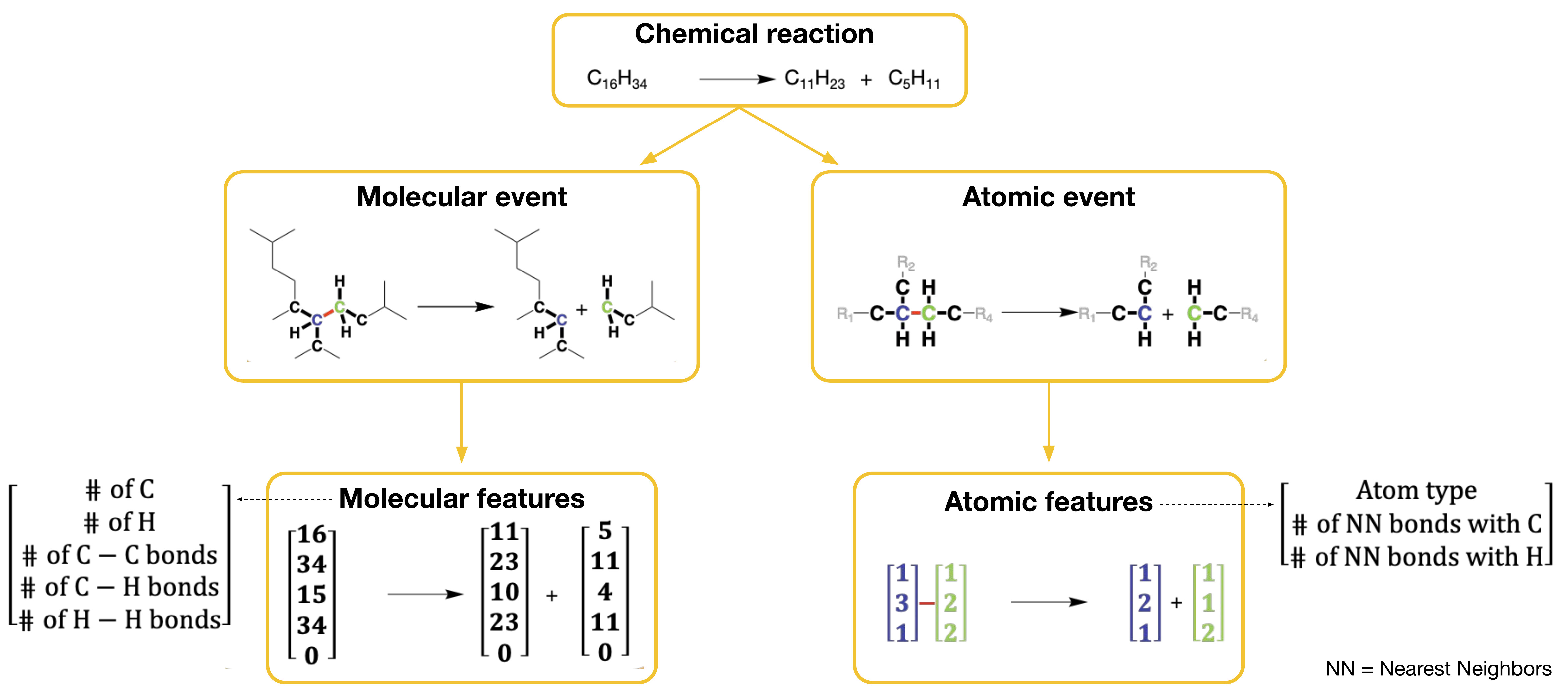}
  \caption{Illustration of how the same chemical reaction is numerically characterized by the two different types of chemical representations considered in this manuscript. The ``molecular features'' involve assigning a numerical fingerprint to each molecule and representing a chemical reaction by the interaction between these molecular fingerprints. Molecular features are chemically intuitive and most commonly. Here we introduce a novel representation that we refer to as ``atomic features''. This representation employs a more local description were each atom has its own set of features and chemical reactions are described locally by the interaction between the atomic fingerprints. Atomic features provide a smaller features space than molecular features, can be parameterized with smaller sets of MD data, and they also enable the determination of the chemical reactivity of systems containing novel chemical species not observed before.}
  \label{fig:representations}
\end{figure*}

Here, we propose an alternative strategy for the extraction of reaction mechanisms and reaction rates from MD simulations, namely the employment of atomic-level features (Fig.~\ref{fig:representations}). Describing reactions at the atomic level comes naturally from the fact that most of the MD potentials are characterizing interactions at the atomic scale rather than at the molecular level. This idea dates back to the work of Stillinger and Weber \cite{stillinger1985computer} and others in the 1980s upon developing of interatomic potential forms beyond simple pair potentials. This novel strategy naturally leads to the parametrization of KMC models that can not only extend the time scale of MD simulations but also predict the mechanisms and rates of reactions never observed in MD simulations. Moreover, we demonstrate that our approach results in a much more compact description of chemical kinetics of hydrocarbons, requiring less data from costly atomistic simulations in order to train models that are just as effective as previous approaches.

\section{Methods}
In this section we present the approach employed to parametrize a KMC model using atomic and molecular features. Reaction rates and reaction mechanisms are extracted directly from MD simulations. The general framework is summarized in Figs.~\ref{fig:simulation_approach} and \ref{fig:representations}.

\subsection{Molecular Dynamics simulations.} 
The MD simulations were performed using LAMMPS \cite{PLIMPTON19951,LAMMPS_website} and the ReaxFF potential \cite{van2001reaxff,ReaxFF_2, AKTULGA2012245} with parameters as described by \citet{PhysRevB.81.054103}. Independent simulations were run starting with either 160 molecules of methane (CH$_4$), 125 molecules of ethane (C$_2$H$_6$), 64 molecules of isobutane (C$_4$H$_{10}$), or 64 molecules of octane (C$_8$H$_{18}$). Temperature and pressure were increased together to 3,300$\,$K and from 1,013$\,$hPa to 40.53$\,$GPa using a Nosé-Hoover chain thermostat and barostat with chain length 3 and damping parameter of 2.4$\,$fs for temperature and 60$\,$fs for pressure.\cite{NoseHoover1, NoseHoover2, NoseHoover3, NoseHoover4, NoseHoover5} The ramping process was spread over 24$\,$ps with a timestep duration of 0.12$\,$fs. Finally, the system was kept at 3,300$\,$K and 40.53$\,$GPa for 500$\,$ps using the same thermostat, barostat (but now with damping parameter of 14.4$\,$fs), and timestep of 0.12$\,$fs. During this 500$\,$ps period the atom coordinates were saved every 12$\,$fs in order to perform an analysis of the system's chemical reactivity. These conditions of temperature and pressure are chosen because they are considered as the approximate thermodynamic conditions of gas-giant planetary interiors \cite{kraus2017formation}, where it is speculated that a rich hydrocarbon chemistry might be present and chemical kinetic evolution of solid phases of carbon could impact internal planetary dynamics.

\subsection{Bond analysis.} 
In order to determine chemical reactivity it is necessary to capture the formation and breaking of chemical bonds in the MD simulations. Here, this is achieved by using the following criteria: two atoms are considered bonded if they are separated by less than a bond cutoff length $\lambda$ for longer than a time period $\tau$. Similarly, a bond between two atoms is considered to have been broken if two atoms initially bonded are separated by a distance larger than $\lambda$ for a period of time longer than $\tau$. The values for $\lambda$ and $\tau$ were respectively taken from Refs.~\onlinecite{bondlengthcriterion} and ~\onlinecite{QianArticle}, where a careful analysis lead to the optimal values of $\lambda = 1.98\,\si{\angstrom}$ for C-C bonds, $\lambda = 1.57\,\si{\angstrom}$ for C-H bonds, $\lambda = 1.09\,\si{\angstrom}$ for H-H bonds, and $\tau = 0.096\,$ps. In Ref.~\onlinecite{bondlengthcriterion}, the bond length criteria for a bond were chosen to be the first minimum of the radial distribution function. In Ref.~\onlinecite{QianArticle}, the bond duration criterion was optimized to obtain the lowest error between the predictions of the KMC model and the real MD trajectories. Other construction of a bond from MD data can be found in the literature such as recent work by Rice et al. \cite{rice2020heuristics}, where bonds are defined using bond distance and vibrational criteria. A more elaborate comparison between the two different bond definitions is performed in Ref.~\onlinecite{rice2020heuristics}.

\subsection{Reaction representation.} 
Two different representations of chemical reactions are considered here (Fig.~\ref{fig:representations}), each one leading to a different set of numerical features characterizing a reaction. The first representation is a chemically intuitive one: each molecule has a numerical fingerprint consisting of features that count the number of each chemical element in the molecule as well as the number of bonds between each pair of elements. Whenever a reaction occurs the quantities registered are the types of molecular fingerprints involved. Because of this we refer to this representation as ``molecular features''. This is a well-known representation in the literature, e.g., Refs.~\onlinecite{QianArticle,EnzeArticle}.

In this article we introduce a second type of representation for chemical reactions in which the characterization occurs more locally, at the atomic level. In this representation each atom has its own numerical fingerprint (Fig.~\ref{fig:representations}) consisting of features that identify the chemical element of the atom and the number of bonds formed with each chemical element available. Whenever a reaction occurs the quantities registered are only the types of atomic fingerprints involved. Because of this we refer to this representation as ``atomic features''. While molecular features can lead to reactions involving many molecules (resulting in many bonds being simultaneously broken or created), we assume that the atomic features always involve only the breaking or formation of a single bond.

\subsection{Reaction rates estimation.} 
Over the course of the MD simulations all reactions observed were recorded and their numerical fingerprints computed using both representations, atomic and molecular. Here we describe how this information was employed to estimate reaction rates. Our approach follows the work of \citet{QianArticle}.

The state of the system at time $t$ is represented by a vector of concentrations $\mathbf{X}(t)$. For the molecular features each component is the concentration of one of the molecular species (i.e., molecular fingerprint), while for atomic features each component is the concentration of one atomic fingerprint. The probability of occurrence of a reaction $j$ in the time interval $[t,t+\Delta t]$ is $a_j(\mathbf{X}(t)) \Delta t$, where $a_j$ is known as the propensity function. Notice that a reaction $j$ is considered to be a molecular reaction for the molecular features, while for atomic features a reaction $j$ is a bond breaking or formation event. The propensity function is $a_j(\mathbf{X}(t)) = k_j h_j(\textbf{X}(t))$, where $k_j$ is the reaction rate coefficient and $h_j(\mathbf{X}(t))$ is the combinatorial number of times that reaction $j$ could have taken place given the system state $\mathbf{X}(t)$. For atomic features we have $h_j(\mathbf{X}(t)) = X_m(t)$ for bond breaking,  $h_j(\mathbf{X}(t)) = X_m(t)X_{m'}(t)$ for bond formation between two different atomic fingerprints, and  $h_j(\mathbf{X}(t)) = X_m(t)(X_m(t) -1)$ for bond formation between identical atomic fingerprints. For molecular features $h_j(\mathbf{X}(t))$ has a similar form but more than two reactants might be involved, in which case the same combinatorial argument can be applied (see Ref.~\onlinecite{QianArticle} for more details).

The calculation of $k_j$ is more intricate and requires the following assumptions. First, the time interval $\Delta t$ is assumed to be short enough for the propensity function $a_j(\mathbf{X}(t))$ to be considered constant during that time interval. Second, the number of times $n_j(t, t+\Delta t)$ that reaction $j$ occurs in the time interval $[t,t+\Delta t]$ is assumed to follow a Poisson distribution with parameter $a_j(\mathbf{X}(t)) \Delta t$. Finally, the Poisson random variables of all reactions are assumed to be conditionally independent given $\mathbf{X}(t)$. With these assumptions it becomes possible to use maximum-likelihood estimation to calculate the reaction rate coefficient $k_j$ as
\begin{equation}
   k_j = \frac{\sum_t n_j(t, t+\Delta t)}{\Delta t \sum_t h_j(\mathbf{X}(t))} = \frac{N_j}{\Delta t H_j},
   \label{eq:reaction_rate}
\end{equation}
where $N_j = \sum_{t} n_j(t, t+\Delta t)$ is the total number of times that reaction $j$ occurred and $H_j = \sum_t h_j(\mathbf{X}(t))$ is the total number of times reaction $j$ could have occurred. The 95\% confidence interval of $k_j$ can be calculated using the Fisher information of the likelihood \cite{rice2006mathematical}. Few lines of calculations described in Ref.~\onlinecite{rice2006mathematical} gives a 95\% confidence interval of: 
\begin{equation}
    k_j \pm  1.96\sqrt{\frac{k_j}{\Delta t H_j}}.
    \label{eq:std_error}
\end{equation}
Yet, reactions have rates that can vary by orders of magnitude. Thus, it is often useful to normalize the size of the 95\% confidence interval of Eq.~\eqref{eq:std_error} by $k_j$ when comparing the accuracy of different reaction rates, leading us to the normalized size of the 95\% confidence interval (NSCI):
\begin{equation}
    \text{NSCI}(k_j) =  \frac{2\times1.96\sqrt{\frac{k_j}{\Delta t H_j}}.}{k_j} = 3.92\sqrt{\frac{1}{N_j}}.
    \label{eq:normalized_std_error}
\end{equation}

\subsection{Kinetic Monte Carlo.\label{subsec:KMC}} 
Once the set of all possible reactions $j$ and reaction rates $k_j$ have been obtained from the MD simulations it is possible to reproduce the system time evolution using a Kinetic Monte Carlo (KMC) approach known as the Gillespie stochastic simulation algorithm \cite{gillespie1976general,higham2008modeling}, which we briefly review next. Given the state of the system at time $t_0$, $\mathbf{X}(t_0)$, the KMC algorithm determines the state $\mathbf{X}(t_1)$ at a future time $t_1$ by selecting a single reaction to occur between $t_0$ and $t_1$. The time $t_1 = t_0 + \tau$ at which the next reaction occurs is randomly selected from an exponential distribution $p(\tau | \mathbf{X}(t_0)) = a(t_0) \exp[-a(t_0) \tau]$, where $a(t_0) = \sum_i a_i(\mathbf{X}(t_0))$. The reaction taking place at $t_1$ is also selected randomly, with reaction $j$ being selected with probability $p_j(t_0) = a_j(\mathbf{X}(t_0)) / a(t_0)$. Applying the modifications caused by reaction $j$ to $\mathbf{X}(t_0)$ results in $\mathbf{X}(t_1)$. 

Atomic and molecular features are both capable of describing the same set of chemical reactions. Yet, there is a fundamental difference between them in how the state of the system evolves in time during a KMC simulation. For molecular features the state of the system $\mathbf{X}(t)$ is simply the number of each distinct molecular fingerprint currently present in the system. The set of all possible molecular fingerprints (i.e. the length of vector $\mathbf{X}$) is predetermined by those fingerprints observed in an MD simulation. Thus, by using molecular features the KMC simulation is constrained to never exhibit any reaction event or molecular species that has not been observed in the MD simulation. Such is not the case for the atomic features, where the state of the system $\mathbf{X}(t)$ is composed of the number of distinct atomic fingerprints currently in the system. When a reaction is chosen using the KMC algorithm, it is necessary to randomly select the pair of atoms participating in this reaction. Each pair of atoms with the correct atomic fingerprints (i.e. the fingerprints involved in the reaction) has the same probability to be chosen. Once the two atoms are selected, a bond between them is created in the case of a bond creation, or is broken in the case of a bond breaking. To keep track of the connectivity between the different atoms an adjacency matrix is employed. The adjacency matrix is a square matrix with a number of rows and columns equal to the number of atoms in the system. The elements of this matrix are equal to 1 when the pair of atoms is connected and 0 otherwise. In the case of an MD simulation, this adjacency matrix is initialized at time 0 only using the bond distance criterion. For example, if the distance between two carbons is less than $1.98\,\si{\angstrom}$ they are going to be considered as bonded. The adjacency matrix is then updated using the bond distance and duration criterion at each timestep, as discussed earlier. In the case of a KMC simulation, the adjacency matrix is initialized using the initial adjacency matrix of the MD simulation of the same system. The adjacency matrix is then updated using the reactions produced during the KMC simulation steps. Notice that over the course of the KMC simulation the adjacency matrix contains all the information necessary in order to compute the atomic features for each atom in the system. This matrix also allows us to reconstruct the network of connections between the atoms at each time step, i.e., it allows us to define the molecules present in the system from the atomic fingerprints. It is during this reconstruction step that the atomic features can result in molecular species that have never been observed in the MD simulation. Note that a given adjacency matrix, obtained from the atomic features, reconstructs a unique set of molecules described by the molecular features, however a set of molecular features is not associated with a unique adjacency matrix. Moreover, the molecular features do not differentiate between certain isomers of the same molecule, whereas the adjacency matrix reconstructs a specific isomer of a molecule.

In a KMC simulation, when a reaction is picked, it is assumed here that all species with the correct fingerprint are equally likely to react. But this assumption may not always hold for the atomic features. For example, when a very long carbon chain is present in the system, atoms in the long chain may not all behave in the same way. Long carbon chains tend to contract themselves into large particles that are the result of incomplete hydrocarbon pyrolysis. When that happens atoms in the periphery of the particle may react with the remaining of the system, while atoms deep inside the particle would more likely react with atoms of the large particle they belong to (Fig. S1). In such cases, the assumption that atoms with the correct atomic fingerprints are all equally likely to react could be broken and this problem can result in some limitations of the atomic features. For example, in the MD simulation, atoms on the periphery of the large particle could react with atoms out of this particle which would result in the growth of the particle, whereas atoms deep inside the particle would react with atoms of the same particle, which would not result in the growth of the particle. Yet, in the KMC simulation atoms on the periphery or deep inside the particle would have the same probability to react with atoms out of or inside the particle, which would disrupt the growth of large particle. In order to avoid this, the growth of large carbon chains was tracked and the chemical reactivity analysis halted at the simulation time when the longest molecule in the system contained 10$\,$\% of all the carbons in the system. This constraint resulted in considering only the first 300$\,$ps of the CH$_4$ simulations, 100$\,$ps of the C$_2$H$_6$ simulations, 50$\,$ps of the C$_4$H$_{10}$ simulations, and 50$\,$ps of the C$_8$H$_{18}$ simulations. The different time needed to reach 10$\,$\% reflects the fact that systems with larger carbon content and larger initial molecules result in faster growth of carbon chains.

\subsection{Error calculation.} 
An error metric is needed in order to compare the system time evolution predicted by KMC to the results of the MD simulation. An appropriate option is to measure and compare the concentration of the most numerous molecules: CH$_4$, C$_2$H$_6$, and H$_2$. Another good indicator of the accuracy of KMC simulations is the number of carbon atoms in the longest molecule, as the growth of long carbon chains is also a function of the system kinetics. Tracking the size of the longest carbon chain is a way to show that the models can follow ‘rare’ species since the longest carbon chain only occurs in small quantities (rarely more than one). Following other 'rare' species, such as unstable ones, is difficult because their number at each timestep is either 0 or 1 and interesting statistics cannot be extracted in this case. Thus, in this article we often compare the time evolution of CH$_4$, C$_2$H$_6$, H$_2$, and number of carbons in the longest molecule as predicted by KMC and MD. In order to increase the statistical accuracy of the comparison the MD simulations results are averaged over three independent simulations while the KMC simulations results are averaged over 20 independent simulations.

Besides comparing the time evolution, it is also useful to have a more condensed and objective metric that summarizes the errors accumulated over the entire time-evolution trajectory. For that purpose we define the following quantity
\begin{equation}
  \text{Error} = \frac{1}{T} \sum_n \sum_t \frac{\left| \mu_n^\text{MD}(t) -          \mu_n^\text{KMC}(t)\right|}{\sigma_n^\text{MD}(t)},
  \label{eq:error}
\end{equation}
where  $n$ is one of the four species of interest mentioned above, $T$ is the number of timesteps, $\mu_n^\text{MD/KMC}(t)$ is the number of species $n$ at time $t$ averaged over all independent simulations, $\sigma^\text{MD}_n(t)$ is the standard deviation of the number of species $n$ at time $t$ between three independent MD simulations of the same system. The division by the standard deviation serves to account for the variability of MD results when evaluating the discrepancy between MD and KMC. Consequently, time intervals in which MD results present large variance influence the error calculation less strongly.

\section{Results\label{sec:results}}
\begin{figure*}[htb]
  \centering
  \subfloat[][]{
    \label{fig:transferability_atomic}
    \includegraphics[width=.9\linewidth]{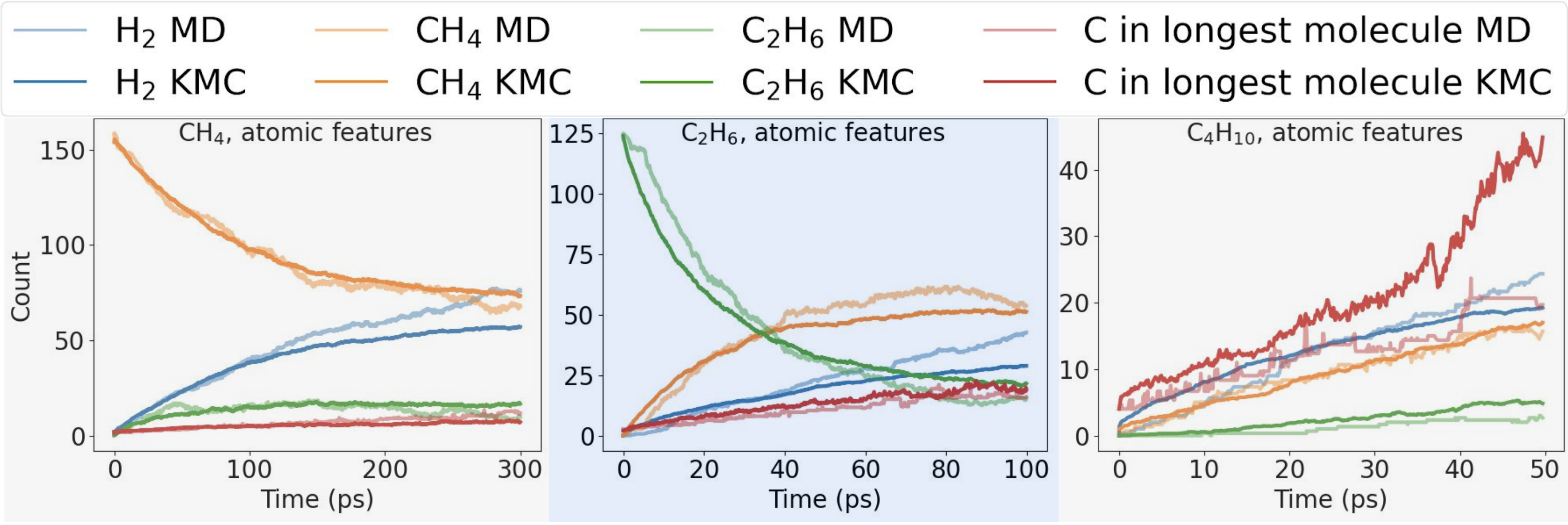}
  } \hfill
  \subfloat[][]{
    \label{fig:transferability_molecular}
    \includegraphics[width=.9\linewidth]{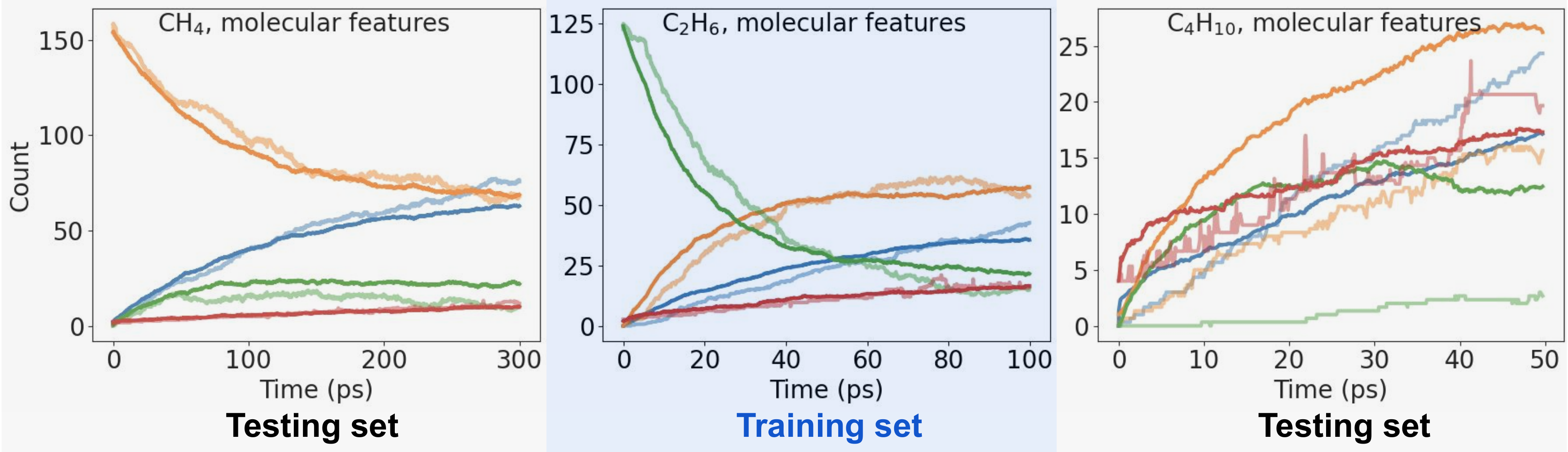}
  } \hfill
  \subfloat[][]{
    \label{fig:transferability_error}
    \includegraphics[width=.6\linewidth]{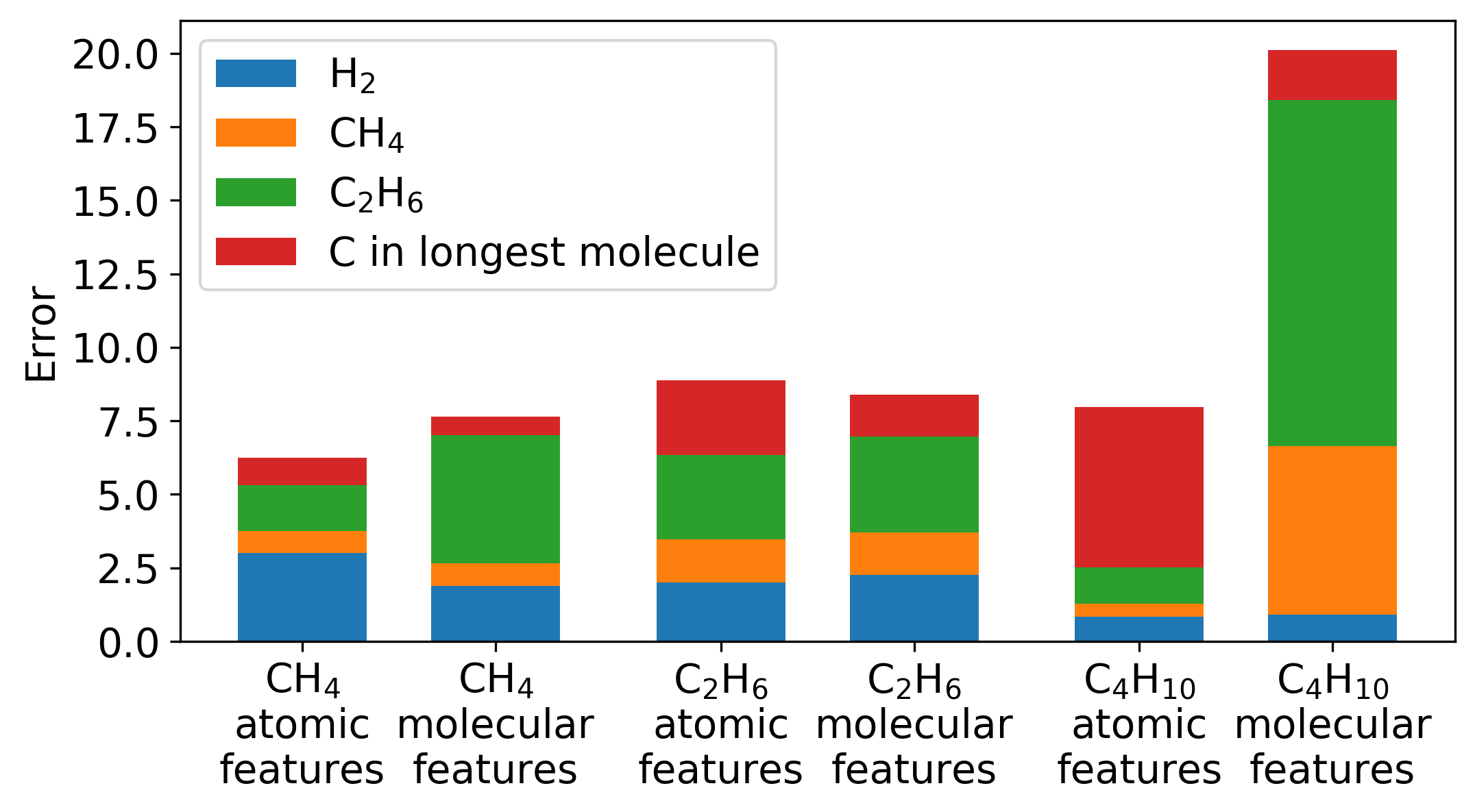}
  }
  \caption{Time evolution of the three most numerous molecules (H$_2$, CH$_4$, and C$_2$H$_6$) and the number of carbons in the longest molecule. The KMC results are obtained using (a) atomic features and (b) molecular features. The simulations initial state contained only CH$_4$ (left), C$_2$H$_6$ (middle), or C$_4$H$_{10}$ (right) molecules. The KMC simulations were parametrized using a single MD simulation that started with only C$_2$H$_6$ molecules. Therefore the middle plots in (a) and (b) show how well the KMC reproduced the simulation it was parameterized with (training set in blue), and the left and right plots in (a) and (b) show transferability of the KMC model to initial compositions it was not trained on (testing sets in grey).  Results for the MD simulations represent the average of three independent simulations, while KMC results are the average of 20 independent simulations. (c) Total trajectory error computed according to Eq.~\eqref{eq:error}. Atomic and molecular features reproduce well the results of the MD simulation for which they were trained on (i.e., starting with only C$_2$H$_6$) and are both equally transferable to MD simulations starting with only CH$_4$, which has a different C/H ratio. Atomic features result in a more transferable KMC model for a system starting with C$_4$H$_{10}$ molecules, especially for the kinetics of small molecules. Molecular features are less transferable (i.e., larger total error), but better reproduce the time evolution of the number of carbon in the longest molecule.}
  \label{fig:transferability}
\end{figure*}

Reactions and their rates, Eq.~\eqref{eq:reaction_rate}, were computed for atomic and molecular features employing the entire trajectory of a single MD simulation of a system starting with only C$_2$H$_6$ molecules. Using this set of reactions a KMC model was parametrized and KMC simulations were run to study the chemical kinetics of a system with the exact same starting configuration as the MD simulation (i.e. same amount of C$_2$H$_6$ molecules). In Fig.~\ref{fig:transferability} we compare the time evolution of the system according to both simulation methods for the atomic and molecular features. It is visually clear that KMC simulations with either type of features are able to reproduce the results of the more computationally expensive MD simulations. The metric of Eq.~\eqref{eq:error}, shown in Fig.~\ref{fig:transferability_error}, confirms these observations: atomic and molecular features present similar total error accumulated over the entire trajectory. The metric is normalized by the standard deviations between three independent MD simulations which are not shown here for readability purposes but can be found in Fig. S2. Despite the similarity of the results there is a large difference in the number of unique reactions observed: while the molecular features result in $845$ unique molecular reactions the atomic features produce only $122$ unique atomic reactions. The total number of reactions observed is $2,683$ for the molecular features and $3,358$ for the atomic features. These two numbers are different because a reaction with the atomic features can only be one bond breaking or creation, whereas there is no such constraint with the molecular features. Equation \eqref{eq:normalized_std_error} shows that the accurate estimation of rate $k_j$ requires reaction $j$ to occur many times. Thus, the atomic features lead to a more accurate and compact KMC model representation of the atomistic MD results.

In terms of computation costs, a single MD simulation takes around one full day to run in parallel on 40 CPUs. Meanwhile, the feature extraction process for either type of features takes only two minutes a single CPU and a KMC simulation running in two minutes in a single CPU. This represents a speedup on the order of $14,000$ in terms of CPU-hours.

\subsection{Model transferability.} 
Next, we test whether the set of reactions learned from a single MD simulation starting with C$_2$H$_6$ is capable of reproducing the kinetics of systems with different ratios of carbon to hydrogen. The MD simulations starting from CH$_4$ molecules or C$_4$H$_{10}$ represent two test cases where the C/H ratio is above and below, respectively, that for the C$_2$H$_6$ starting condition. Figure \ref{fig:transferability} shows that KMC simulations with both types of features perform similarly well in reproducing the CH$_4$ system kinetics, with the atomic features having a lower total error than the molecular features. Atomic features do seem to have a relatively larger error for the H$_2$ time evolution, while molecular features have a similarly larger error for the C$_2$H$_6$ time evolution. The scenario is different for the reproduction of the MD simulation starting with C$_4$H$_{10}$. Now, atomic features result in a much lower total error (by a factor of $2.5$) than molecular features. Yet, it is noticeable that the majority of the error for atomic features stem from the reproduction of the size of the longest carbon chain. Further analysis of this discrepancy is postponed to Section~\ref{sec:discussion}.

\subsection{Time extrapolation.}
\begin{figure*}[htb]
  \centering
  \subfloat[][]{
  \label{fig:time_extrapolation_atomic}
  \includegraphics[width=.9\linewidth]{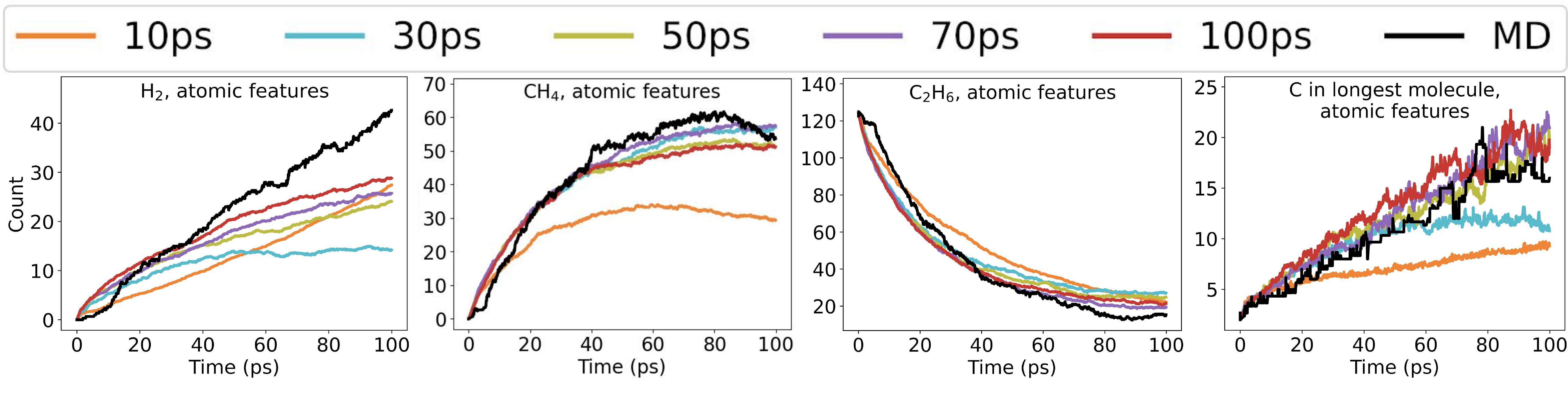}
  } \hfill
  \subfloat[][]{
  \label{fig:time_extrapolation_molecular}
  \includegraphics[width=.9\linewidth]{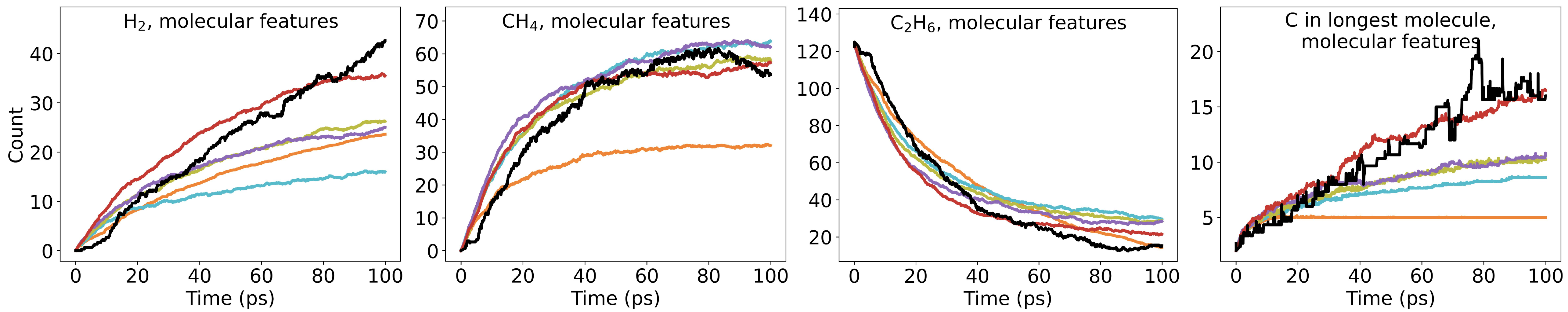}
  } \hfill
  \subfloat[][]{
  \label{fig:time_extrapolation_error_local}
  \includegraphics[width=.485\linewidth]{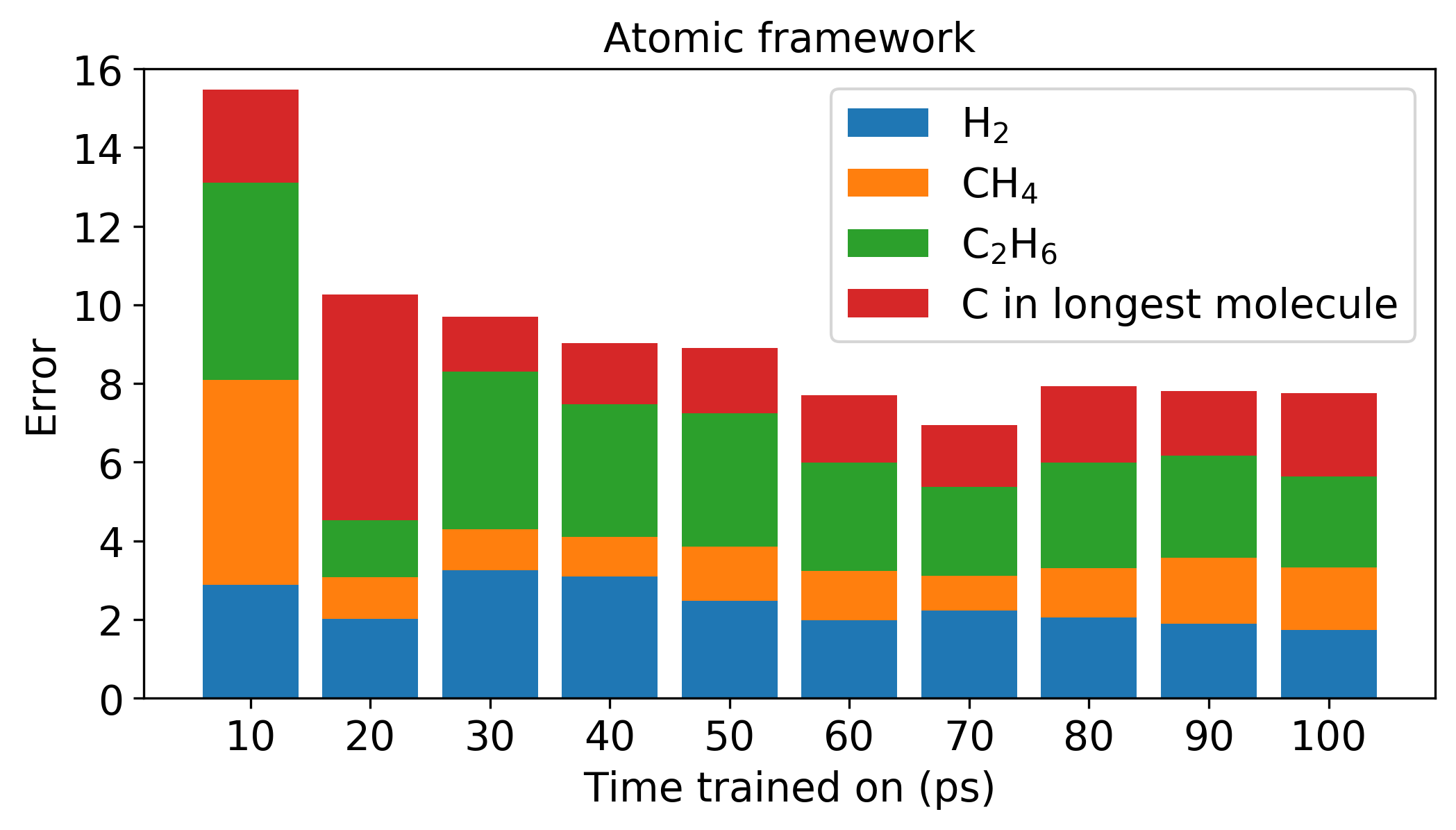}
    }    \hfill
  \subfloat[][]{
  \label{fig:time_extrapolation_error_molecules}
  \includegraphics[width=.485\linewidth]{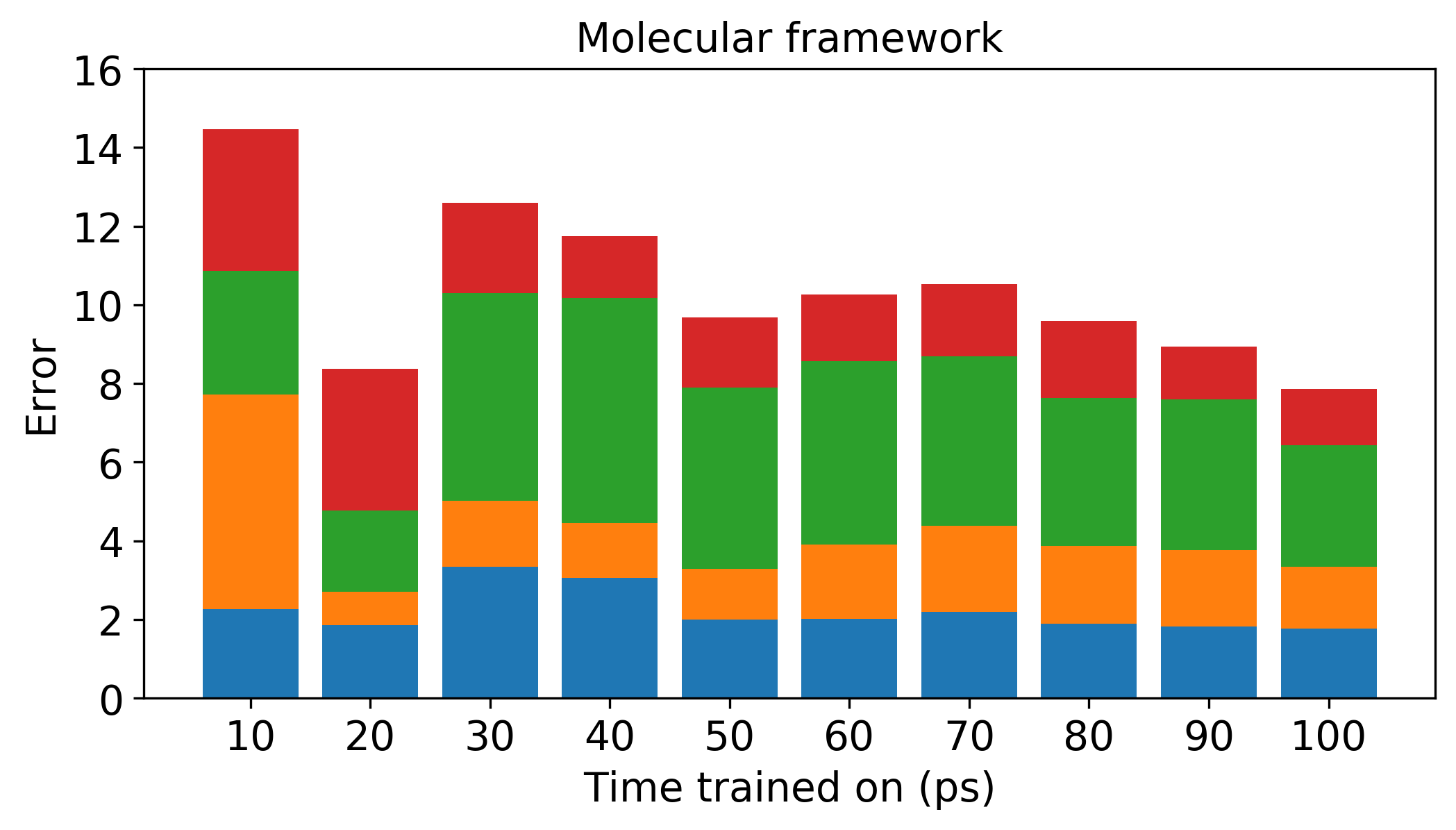}
  }
  \caption{Time extrapolation of MD simulations using KMC with (a) atomic features and (b) molecular features. Time evolution of the three most numerous molecules (H$_2$, CH$_4$, and C$_2$H$_6$) and number of carbons in the longest molecule for simulations with initial state containing only C$_2$H$_6$ molecules. The KMC models were parametrized using only the first 10$\,$ps, 30$\,$ps, 50$\,$ps, 70$\,$ps, and 100$\,$ps of a single 100$\,$ps MD simulation. Results for the MD simulations represent the average of three independent simulations, while KMC results are the average of 20 independent simulations. Total trajectory error of time extrapolation of MD simulations using KMC with (c) atomic features and (d) molecular features. The KMC models were parametrized using only the first 10$\,$ps, 20$\,$ps, 30$\,$ps, 40$\,$ps, 50$\,$ps, 60$\,$ps, 70$\,$ps, 80$\,$ps, 90$\,$ps, and 100$\,$ps of a single 100$\,$ps MD simulation. The trajectory error is computed according to Eq. \ref{eq:error}. The atomic framework learns sufficient information in 60$\,$ps and after that the error fluctuates around an equilibrium value. However, the molecular framework keeps on learning new information and the error keeps on decreasing. Especially, the atomic features are able to reproduce the growth of large carbon chains much faster than molecular features. This is likely due to the fact that the molecular features cannot predict the appearance of molecules it has not observed during its parametrization (i.e. larger carbon chains). Meanwhile, atomic features can estimate the reaction rates of molecular reactions that have not been observed during training by building such molecular reactions from its elementary atomic reaction events.}
  \label{fig:time_extrapolation}
\end{figure*}

One desirable property on KMC simulations is the ability to accurately extrapolate the results of MD simulations to time scales unattainable in MD simulations due to their prohibitive computational costs. In order to compare the ability of atomic and molecular features to perform time extrapolation, KMC models were trained for both types of features using only part of the data extracted from the 100$\,$ps MD simulation of C$_2$H$_6$. In Fig.~\ref{fig:time_extrapolation} it is shown the time evolution of KMC models trained on the first 10$\,$ps, 30$\,$ps, 50$\,$ps, 70$\,$ps, and 100$\,$ps of the MD simulation. The performance of atomic and molecular features is similar, except for the time evolution of the size of the largest carbon chain. It is clear in this case that atomic features present reasonable results when learning from simulations as short as 30$\,$ps, and the error in Fig. \ref{fig:time_extrapolation_error_local} reaches its minimum value after learning from 60$\,$ps simulations and fluctuates around an equilibrium when learning on longer times. However, molecular features only converge to the MD results when training on the entire 100$\,$ps trajectory and the errors in Fig. \ref{fig:time_extrapolation_error_molecules} are decreasing until the model has been trained on the whole simulation.

Atomic features result in a much more compact representation of the chemical reactivity of hydrocarbon systems (122 unique reactions compared to 845 unique reactions for molecular features). Thus, it is reasonable to expect that a KMC model with atomic features can be parametrized with much less data (i.e., shorter MD simulations), which explains in part the capacity that atomic features have shown in Fig.~\ref{fig:time_extrapolation} to reproduce 100$\,$ps of MD simulations of the growth of the largest carbon chain from only 30$\,$ps of data. Another important factor is that molecular features cannot predict the creation of molecules that have not been observed in the MD simulation, limiting its capacity to extrapolate in time the kinetics of growth of large carbon chains. Meanwhile, atomic features can estimate the kinetic rates of reactions that have not been observed during MD simulations by building it from its elementary atomic reaction events.

The fact that both types of features perform similarly for the small molecular species (H$_2$, CH$_4$, and C$_2$H$_6$) is most likely because reactions resulting in the creation or consumption of such small molecules are similarly represented in both types of features, resulting in the same reaction rates. For example, the $\text{H}_2 \rightarrow \text{H} + \text{H}$ chemical reaction has the exact same reaction rate in the atomic representation or molecular representation. It can be observed that the prediction of H$_2$ quickly decreases for both of the models. This can be explained by the low number of molecules H$_2$: there is probably not enough data to obtain accurate estimations of the reaction rates of reactions that are involved in the creation of H$_2$. Indeed, only 40 molecules of H$_2$ are created after 100$\,$ps and this number decreases approximately linearly with time. By comparison, the evolution of the number of CH$_4$ can be predicted accurately after being only trained on 30$\,$ps which also corresponds to having 40 molecules of CH$_4$ in the system.

\section{Discussion\label{sec:discussion}}

\subsection{Mechanism extraction.} 
Atomic and molecular features represent chemical kinetics in different ways. The atomic features framework breaks each molecule into small units composed of one or two atoms and information about their nearest neighbors. These units can be common to different molecules, which allows this framework to capture similarities in reactions involving completely different molecules. Meanwhile, the molecular features framework fundamental unit is the molecule. A reaction is then described as the interactions among these fundamental units generating other fundamental units, without considering the rearrangement of atomic bonds at any step.

There is also a meaningful difference in how chemical kinetics is reproduced through KMC simulations by the two types of features. With atomic features a KMC simulation is capable of creating and consuming molecules never observed in a MD simulation by building their reaction rates from the more elementary atomic reaction events. Molecular features result in KMC simulations that are only able to create and consume molecules that have been previously observed in MD simulations. This difference allows KMC simulations with atomic features to explore a larger variety of chemical reaction pathways when compared to molecular features. Such difference can become important whenever the system trajectory passes through bottlenecks in order to reach different regions of the chemical space. The growth of a large carbon chain can occur in many different ways that can be considered bottlenecks in the chemical trajectory, because each independent simulation only goes through one specific pathway of all possible ones. For example, one can conceive of a trajectory where small molecules such as CH$_4$ are added to a steadily growing chain. This trajectory is much different from one where two independent carbon chains grow to a medium size and then merge to form a large chain. It is evident that a simulation where a chain reaches a determined length can only go through one of these two trajectories.

\begin{figure}[htp]
  \centering
  \subfloat[][]{
  \includegraphics[width=0.45\linewidth]{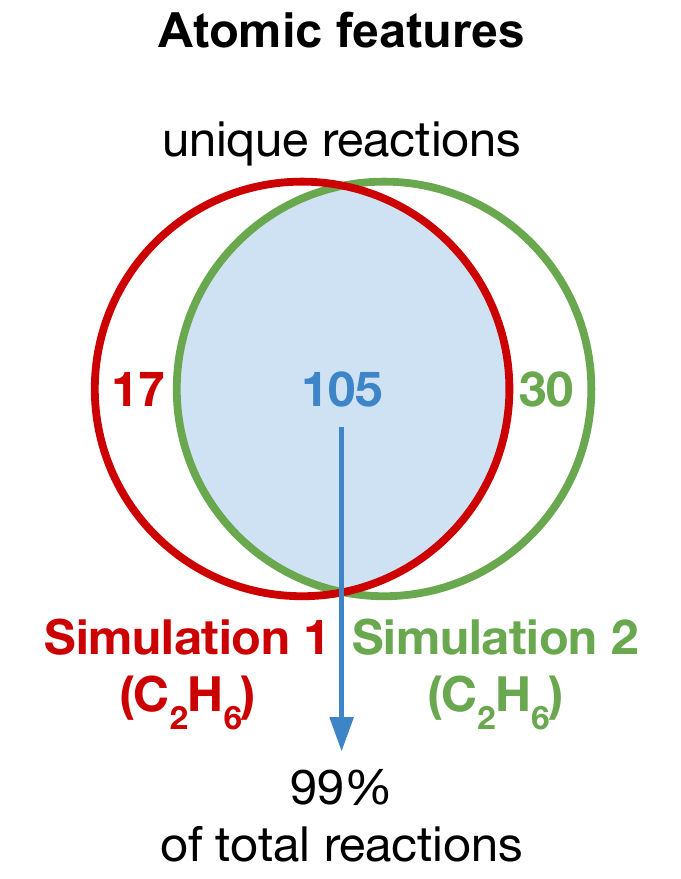}
  \label{fig:Venn_atomic}
  }
  \subfloat[][]{
  \includegraphics[width=0.45\linewidth]{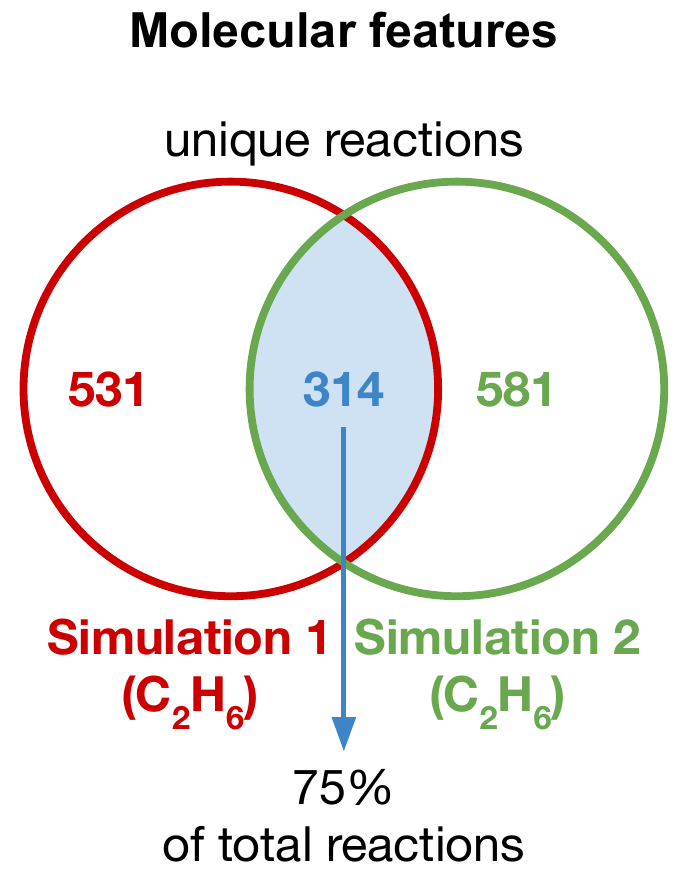}
  \label{fig:Venn_molecular}
  }
  \caption{Unique reactions observed from two independent MD simulations starting from only C$_2$H$_6$ molecules. a) Atomic features result in a more unique and compact representation of the chemical kinetics, with 69$\,$\% of unique reactions (representing 99$\,$\% of total reactions observed) in common among the two MD simulations. b) Molecular features share only 22$\,$\% of reactions among the two MD simulations.}
  \label{fig:Venn}
\end{figure}

In order to offer some evidence of this essential difference between atomic and molecular features we have performed the chemical kinetics analysis of two independent MD simulations with identical initial chemistries (only C$_2$H$_6$ molecules), but the atomic velocities were initialized randomly so that the simulation trajectories would be different. The molecular features resulted in a total of 1,426 unique reactions, with only 314 (22$\,$\%) of those in common among the two identical but independent simulations (Fig. \ref{fig:Venn_molecular}). These 314 reactions in common account for to 75$\,$\% of the total reactions observed, while 68$\,$\% of the total 1,426 unique reactions occurred only once during the entire MD simulation. Meanwhile, atomic features resulted in 153 unique reactions with 105 (69$\,$\%) of them in common among the two identical but independent simulations (Fig. \ref{fig:Venn_atomic}). The 105 unique reactions in common account for 99$\,$\% of the total reactions observed, with only 28$\,$\% of the total 153 unique reactions occurring only once during the entire simulation. Hence, KMC models parametrized using independent MD simulations can be much different when molecular features are employed, while models created with atomic features are essentially identical. 

\begin{figure}[htp]
  \centering
   \subfloat[][]{ 
    \includegraphics[width=0.45\linewidth]{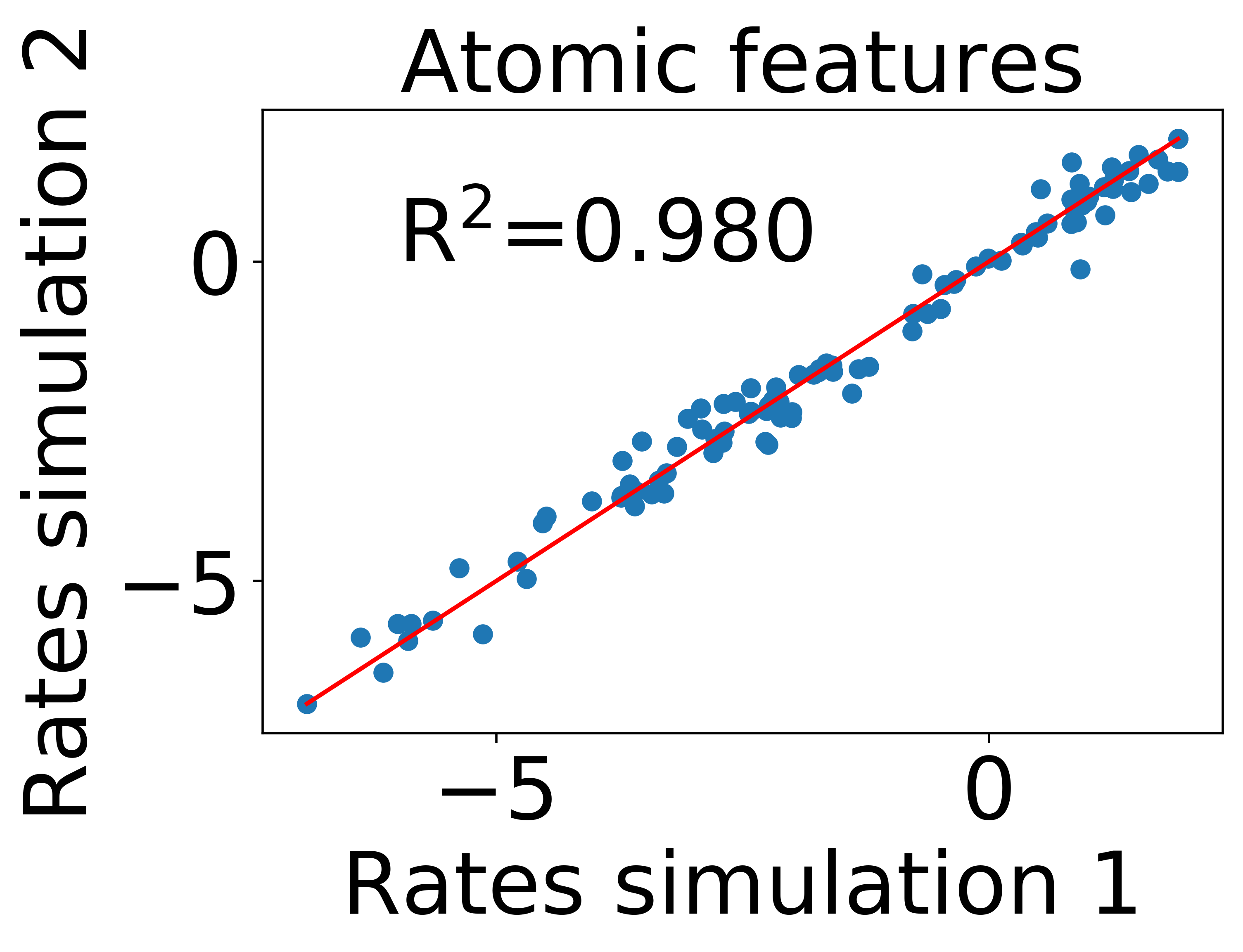}
    \label{fig:loglog_plot_atomic}
  }
   \subfloat[][]{ 
    \includegraphics[width=0.45\linewidth]{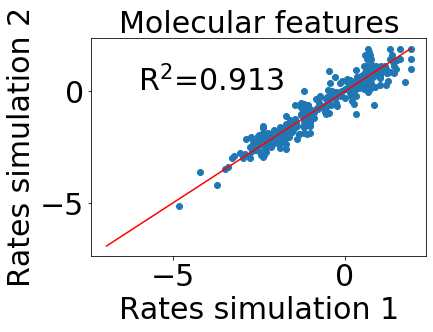}
    \label{fig:loglog_plot_molecular}
  } \hfill
  \caption{Log-scale comparison of the reaction rates in common to two independent MD simulations. The $y=x$ line is shown in red. The coefficient of determination, $R^2$, shows that (a) atomic features result in more similar reaction rates ($R^2 = 0.98$) when compared to (b) molecular features ($R^2 = 0.91$). This is due to the fact that more compact representation of atomic features leads to more statistics collected for each unique reaction.}
  \label{fig:loglog_plot}
\end{figure}

Figure \ref{fig:loglog_plot} compares the rates of reactions for those reactions in common to the two independent MD simulations. The coefficient of determination, $R^2$, shows that atomic features result in more similar reaction rates ($R^2 = 0.98$) when compared to molecular features ($R^2 = 0.91$). In order to achieve an accurate estimation of a reaction rate it is necessary to observe such reaction many times, as shown in Eq.~\eqref{eq:normalized_std_error}. Because atomic features result in a much more compact model (i.e., less unique reactions) the reaction are observed a larger number of times and can be more accurately determined. Indeed, the normalized size of the confidence interval, Eq.~\eqref{eq:normalized_std_error}, is on average $3.14$ for molecular features and $1.96$ for atomic features.

The redundant and lengthy nature of KMC models with molecular features has been acknowledged in the literature before. For example, \citet{QianArticle} and \citet{YanzeArticle} employed techniques such as L1 regularization and computational singular perturbation to reduce the number of unique reactions space by selectively discarding reactions that had small impact on the chemical kinetics. Employing atomic features can be seen as an approach to achieve the same goals without discarding any data collected from MD simulations, consequently making better use of the available data and avoiding any reduction in the accuracy of the reaction rates.

\subsection{Predictions comparison.}
Kinetic models with atomic and molecular features parametrized using a system initiated with C$_2$H$_6$ show excellent transferability to a system initiated with CH$_4$ molecules, but not to C$_4$H$_{10}$. This happens because of different reasons for the different types of features. The atomic features framework overestimates the size of the longest polymer chain, while still performing well on smaller molecules. As discussed in the Sec. \ref{subsec:KMC}, when a long carbon chain grows, the atomic features framework shows its limitations since we suppose it does not incorporate any information that slows down atomic reactivity in large molecules. 

\begin{figure}[htb]
    \centering
    \includegraphics[width=0.48\textwidth]{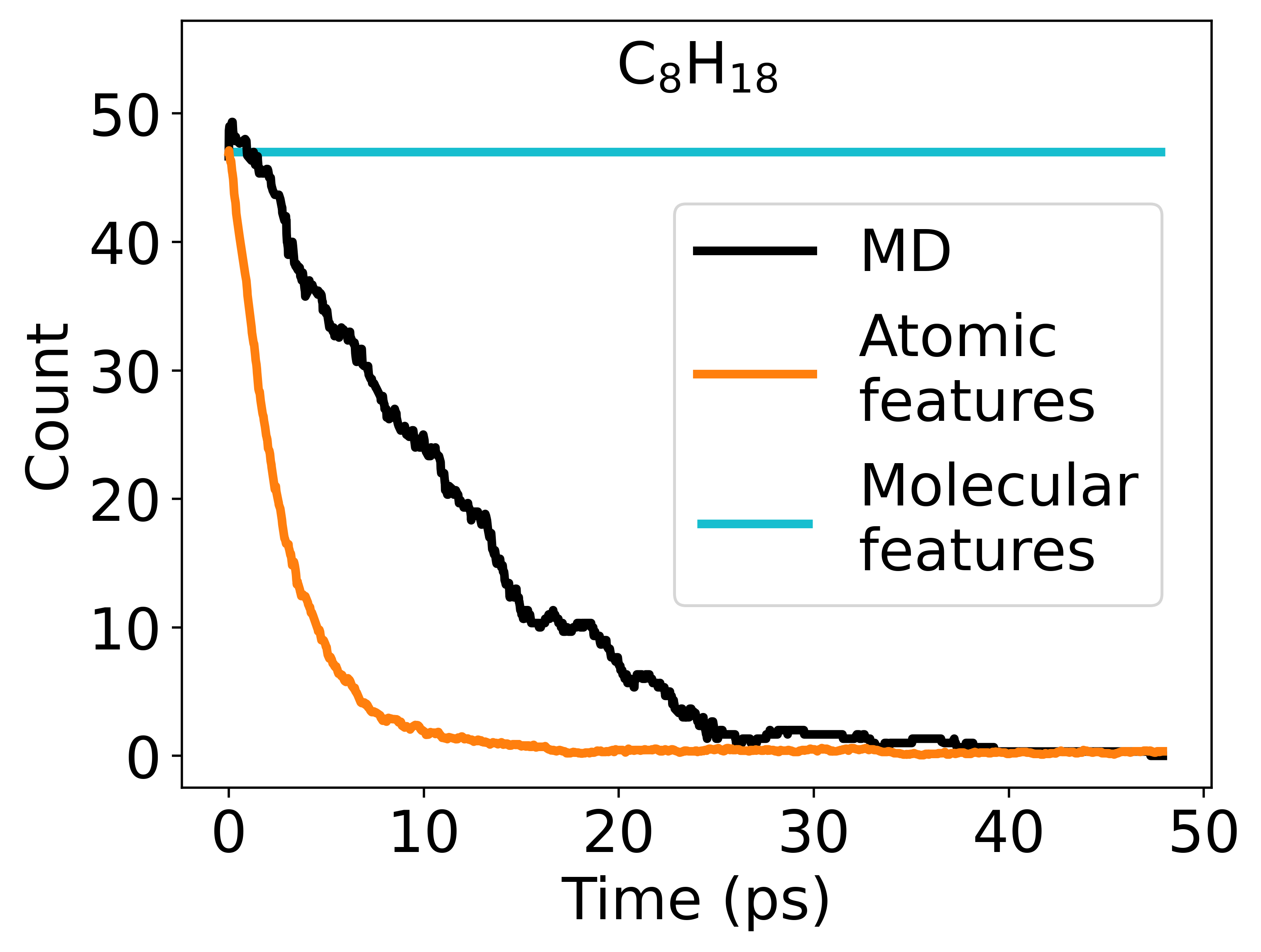}
    \caption{Time evolution of C$_8$H$_{18}$ in a simulation where the initial state contained only C$_8$H$_{18}$. KMC models were parametrized on an MD simulation starting with C$_2$H$_6$ only, where C$_8$H$_{18}$ is never observed. As a consequence, the KMC model using molecular features cannot result in any time evolution for the chemical species. This extreme case illustrates a limitation of employing molecular features. The model employing atomic features model is capable to estimate the reaction mechanisms and reactions rates of C$_8$H$_{18}$ by building it from elementary atomic reactions. Thus, atomic features are still capable of predict chemical kinetics even in such extreme case.}
    \label{fig:octane}
\end{figure}

The limitations in the transferability of the molecular features to a system initiated with C$_4$H$_{10}$ are due to the scarcity of C$_4$H$_{10}$ in the MD simulation that was used to parametrize the KMC model (i.e., starting with C$_2$H$_6$ only). Thus, the molecular features framework has little data on the reactivity of C$_4$H$_{10}$. Parametrizing a KMC model using molecular features can be considered a form of overfitting: the trained model performs well for systems that present only molecules available in the training data set, but it does not extrapolate to molecules it has not seen or that appear in small amounts in the training data set. An extreme case of this lack of transferability can be seen in Fig.~\ref{fig:octane}, where the time evolution of a MD simulation initiated with only C$_8$H$_{18}$ is compared to the predictions of a KMC model parametrized on a MD simulation started with only C$_2$H$_6$ molecules. C$_8$H$_{18}$ was chosen because this molecule never appears in the MD simulation used for training. Thus, the molecular features model has essentially no time evolution as it remains stuck in the initial configuration. Meanwhile, a KMC model employing atomic features =is able to estimate the rate of reactions of C$_8$H$_{18}$ by building it from the rate of atomic level events.

\begin{figure}[htb]
  \centering
  \subfloat[][]{ 
    \label{fig:quick_growth_atomic}
    \includegraphics[width=0.9\linewidth]{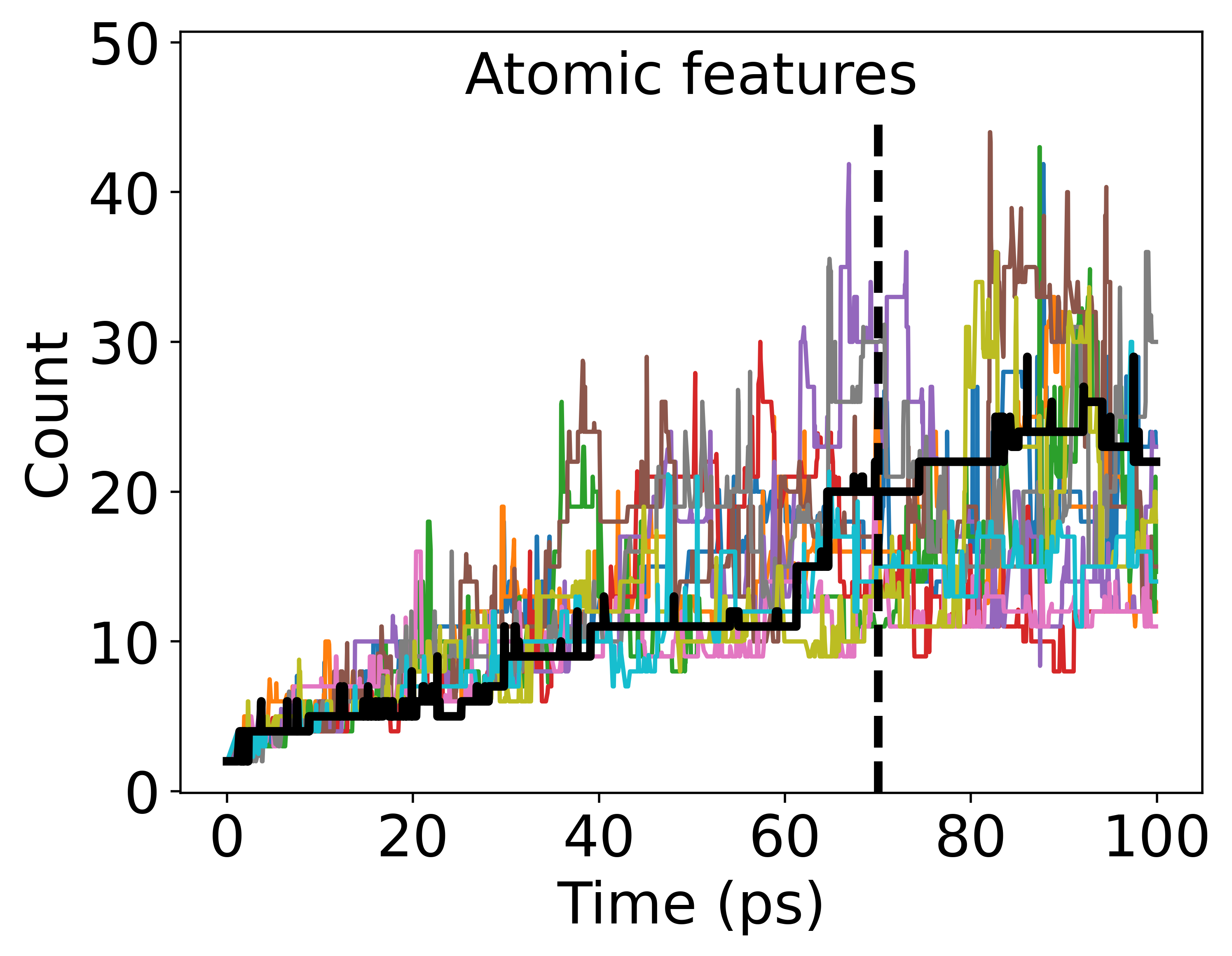}
  } \hfill
  \subfloat[][]{ 
    \label{fig:quick_growth_molecular}
    \includegraphics[width=0.9\linewidth]{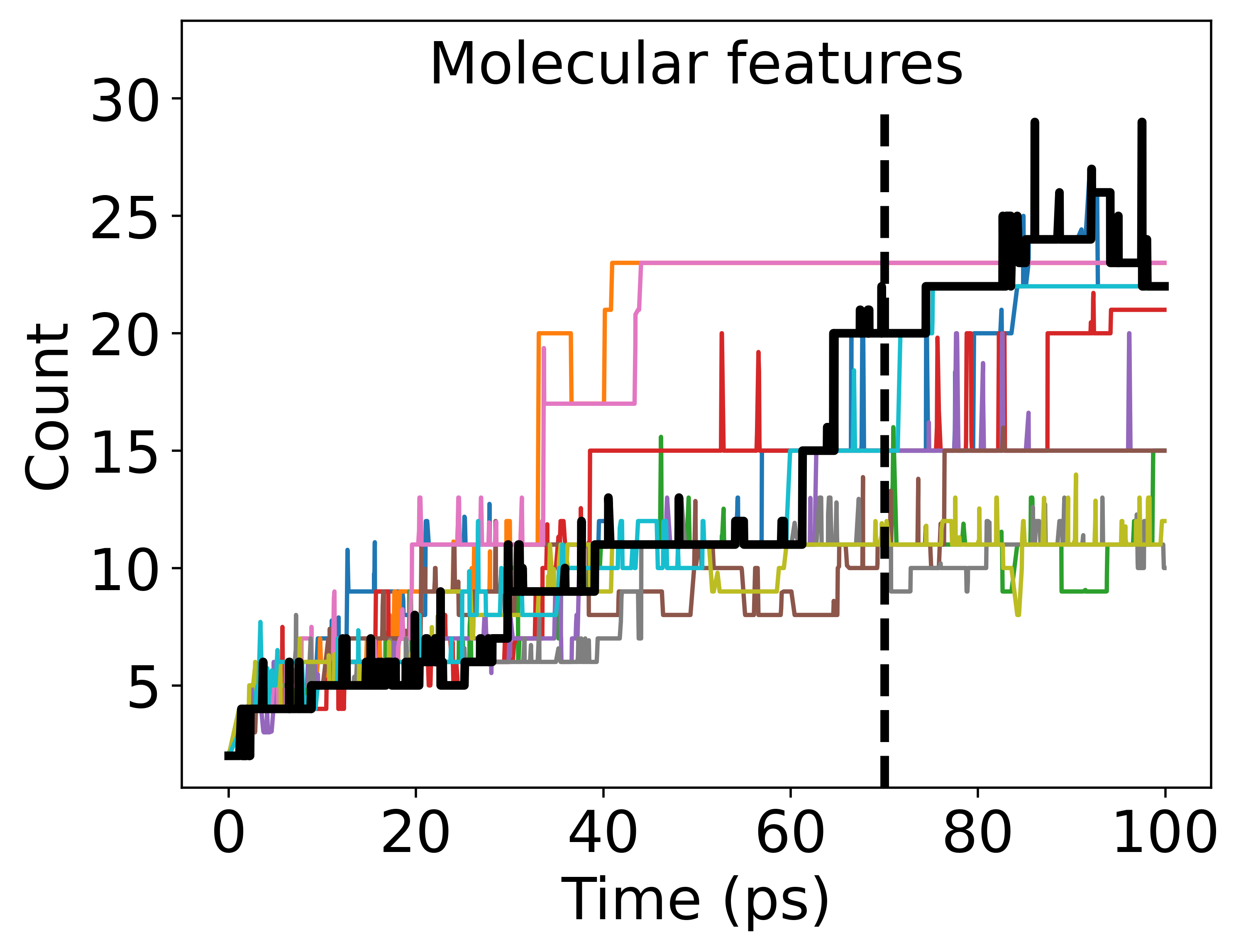}
  } \hfill
  \caption{Time evolution of the number of carbons in the longest molecule. All KMC simulations (shown in colored lines) share the same kinetic model that was parametrized using a single MD simulation (shown in black) initiated with only C$_2$H$_6$ molecules. The KMC curves are trained on 70$\,$ps of this simulation (shown as the dashed black line) with (a) atomic features and (b) molecular features. The two quick changes in size that happen at around 60$\,$ps in the MD simulation are rare events that represent a bottleneck for the growth of the largest carbon chain. KMC models with molecular features can only grow carbon chains past the size of 11 by reproducing these two singular rare events because that is the only known pathway. Thus, if the conditions for the rare events is missed the length of the longest carbon chain in KMC simulations using molecular features will be limited to about 11. It is clear that the atomic features do not suffer from this limitation since they allow the construction of longer carbon chains through multiple pathways not observed in MD simulations. These pathways are found by building the reaction mechanisms and rates from elementary atomic reactions.}
  \label{fig:quick_growth}
\end{figure}

The molecular features framework also presented limitations in time extrapolation to predict the growth of the largest carbon cluster, Fig.~\ref{fig:time_extrapolation}. As discussed in Sec.~\ref{sec:results}, this occurs because KMC simulations employing molecular features are not able to predict the formation of molecules that have not been observed in the simulation it was trained on. As a result, a KMC simulation with molecular features trained on 50$\,$ps of the MD simulation cannot grow a molecule with more than about ten carbons in it. Another consequence of this effect is the scarcity of pathways for the growth of longer chains. Figure \ref{fig:quick_growth} shows that during the MD simulations used to train the KMC model there are two rare events occurring just after 60$\,$ps that cause the chain size quickly increase from 11 to 20 carbons. These events involve the addition of two rare molecules to the longest chain, first a molecule with four carbon atoms and then a molecule with five. This represents a bottleneck for the chemical evolution that can be easily missed by the KMC simulation with molecular features if, for example, all carbon chains in the system grow past the size of five carbons. There are no other pathways learned by the molecular features that increase the size of the system above 11. Therefore, for each MD simulation can only train a KMC model with one or very few pathways of growth past any specific carbon chain size. It is evident in Fig.~\ref{fig:quick_growth} that training a KMC model with atomic features result in no such constraints on the number of pathways leading to the growth of carbon chains past any size. Atomic features allow the kinetic model to build the reaction mechanism and reaction rates of multiply pathways for growth past the size of 11 by estimating them from the elementary atomic events.

\subsection{Atomic features limitations.}
Atomic features also have some specific limitations due to ignoring any information further than the immediate neighbors of reacting atoms. These limitations occur when the atomic features do not describe all of the necessary information to obtain an accurate reaction rate. There are reactions where functional groups that are not immediately adjacent to the reaction site can have a significant effect on the reaction rate. For example, if there is an electron donating or withdrawing group that could stabilize or destabilize an atom and change its reactivity. This situation would occur, for instance, with highly polar bonds, atoms having lone pairs and double or triple bonds. In the system presented here, only carbons and hydrogens are present making the first two examples inconsequential. The last example could have an effect through conjugated systems. In this system, less than 10\% of the carbon-carbon bonds are double or triple bonds and around half of them are in a C$_2$H$_4$ molecule so unable to conjugate, therefore this effect was neglected. Atomic features could also be insufficient in the case where angles between bonds play an important role in the stability of the atoms or of the bonds. For example, in a highly strained cyclopropane, the atomic features would not be able to predict the unstability of this structure. There are less than 10 cycles at a time in the simulations. In order to take into account these limitations, additional features could be added to the atomic features in later work. However, adding additional features would give more unique reactions in the model, which would decrease the number of occurrences of each reaction and so increase the NSCI. More data would then be necessary to train the model to achieve a similar performance.

\subsection{Comparison with elementary reactions}
\begin{figure}[htb]
  \centering
    \includegraphics[width=0.99\linewidth]{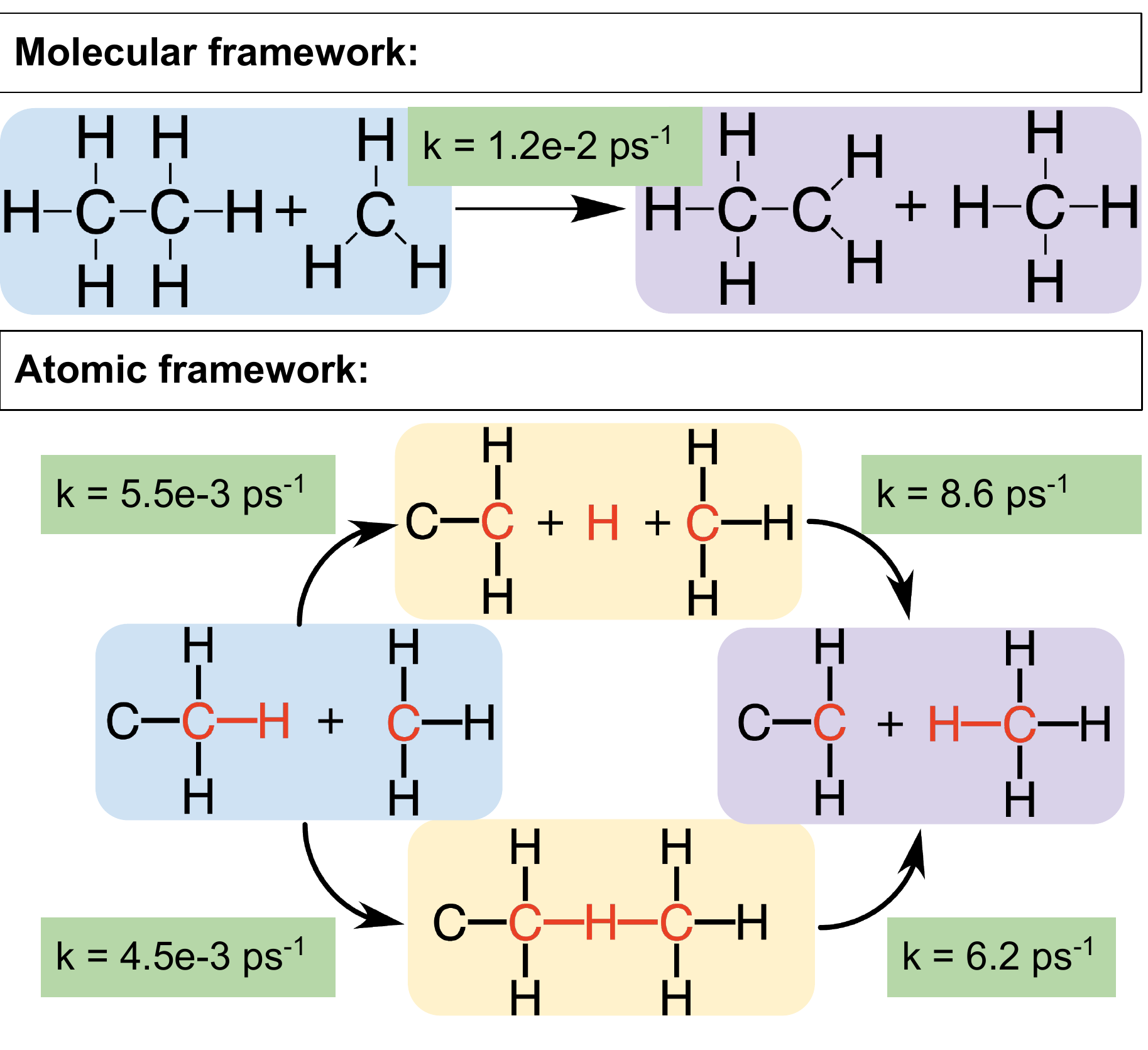}
  \caption{Drawing showing how the same reaction would be represented using the molecular and atomic framework. The molecular framework can be considered as the same as what is usually thought as 'elementary reactions', where one reaction takes into account all of the bond rearrangements occurring during one collision. The atomic framework has the assumption that one reaction is either the breaking or the creation of a bond; therefore an elementary reaction has to be separated in several steps. In addition, these steps can occur in different order, giving several possible paths in the atomic framework for the same molecular reactions. The reaction rates of the atomic framework reactions cannot be exactly related to the reaction rate of the molecular reaction as the two paths consider two steps which involve one or two reactions.}
  \label{fig:reactions}
\end{figure}

Kinetics models are usually built using elementary reactions \cite{smith2011gri, wang2007usc, smith2016foundational}. These reactions are defined as the bond rearrangements that occur during one collision. In the molecular framework, the reactions are elementary reactions; however, in the atomic framework, reactions are defined as only the breaking or the creation of a bond. This definition of the atomic reaction does not in general allow for a unique definition of the elementary rates from the atomic reaction rate. In Fig. \ref{fig:reactions}, it can be observed that one elementary reaction can have several ways of being encoded by the atomic framework, giving rise to a dependence on the populations of intermediate species that the molecular framework rate equation lacks. However, effective molecular framework reaction rates could be obtained by running the kinetic Monte Carlo simulations using the atomic features and resolving the resulting species dynamics into molecular framework rate equations. Such molecular framework rate equations could be useful for integration into other software packages.

It can be noticed in Fig. \ref{fig:reactions} that the atomic framework creates particularly unstable species such as an overcoordinated hydrogen. This exotic species is known to be very unstable by the model and will be quickly consumed, as the high reaction rate of its consumption shows. This approximation allows us to obtain a model with few unique reactions but comes at the cost of observing unrealistic structures for very short times. The molecular framework can also show exotic species, even if reactions can include several bond rearrangements. This usually occurs because the bond length and duration criterion are not perfect to describe bonding. In this case too, unstable species are associated with high reaction rates and their lifetime is very short.

\section{Conclusion}
In this paper it is demonstrated that kinetic models built using atomic features allow the determination of the reaction mechanisms of a complex chemical system, hydrocarbon pyrolysis, and to accurately predict the evolution different systems that rely on the same chemistry using a KMC model. It is shown that atomic features result in more compact kinetic models than molecular features, while being able to predict the appearance of molecules not observed during the parametrization process, which molecular features are not capable of. Atomic features are shown to result in better chemical transferability and time extrapolation due to the ability of kinetic models based on atomic features to explore multiple pathways of chemical evolution by building unknown reaction mechanisms and rates from elementary atomic events. The framework of atomic features considers only the chemical species the reacting atoms and their respective nearest neighbors. This fairly simple description, while powerful, can be easily extended to include other elaborate non-local features. Although our study focused on the specific mechanism of hydrocarbon pyrolysis, the approach developed for the construction of kinetic models can be readily applied to other chemical systems with different levels of complexity.

\section{Supporting Information}
\begin{itemize}
  \item GitHub link to code
  \item Snapshot of the surroundings of some atoms in a long carbon chain
  \item Same results as Figure 3 presented with the standard deviations between the trajectory of 3 MD simulations
\end{itemize}

\section{Acknowledgements}
Material in this paper is based upon work supported by the Air Force Office of Scientific Research under award number FA9550-20-1-0397 and the Department of Energy National Nuclear Security Administration under Award Number DE–NA0002007.

\clearpage
\bibliography{bibliography}

\begin{figure*}[htb]
  \centering
  \includegraphics[width=0.9\textwidth]{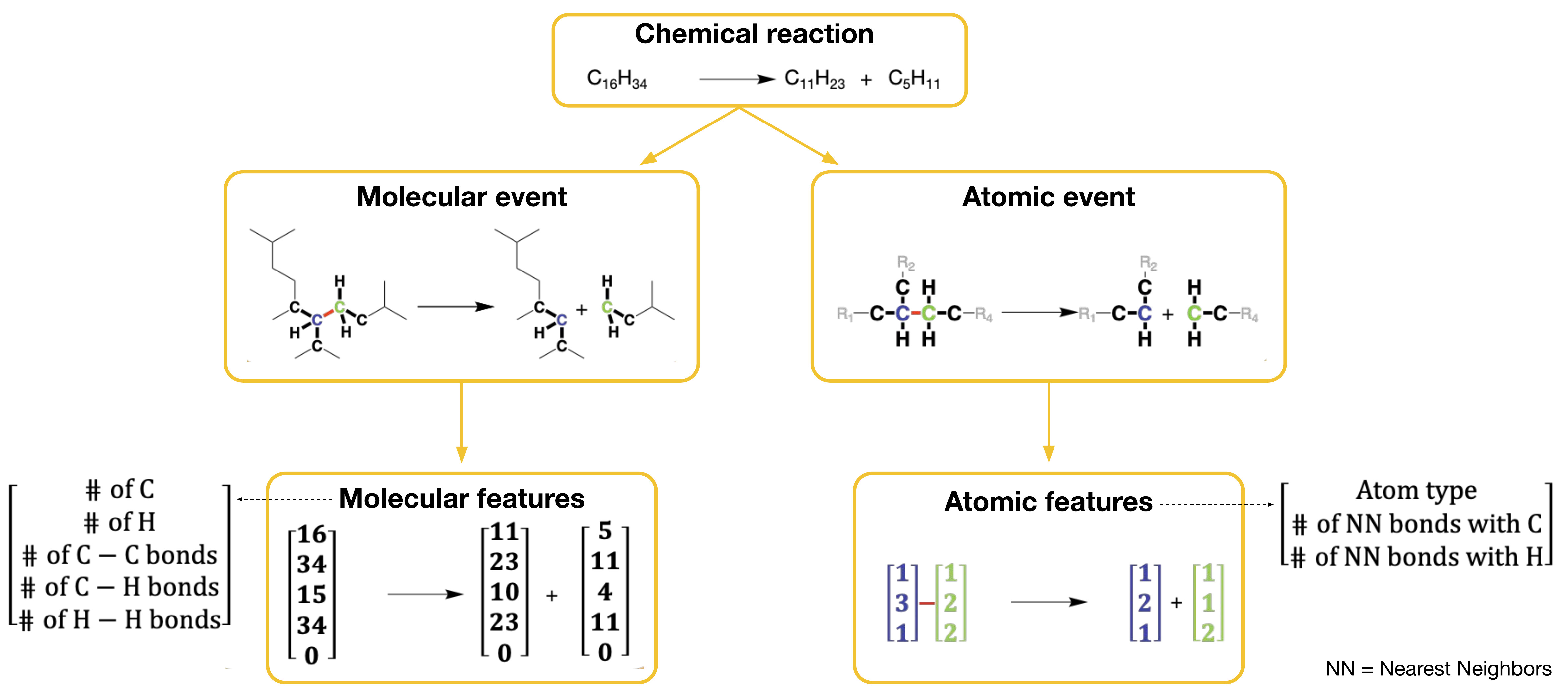}
  \caption{TOC Graphic}
  \label{fig:TOC Graphic}
\end{figure*}

\end{document}